
\def\d#1/d#2{ {\partial #1\over\partial #2} }


\def\pdr{\partial}

\def\al{\alpha}
\def\be{\beta}
\def\ga{\gamma}
\def\Ga{\Gamma}
\def\de{\delta}

\def\eps{\epsilon}
\def\om{\omega}

\def\half{{1\over 2}}
\def\tr{\hbox{tr}}

\def\un#1{\underline{#1}}

\def\L{\bigtriangleup}


\newcount\eqnumber
\def\beq{ \global\advance\eqnumber by 1 $$ }
\def\eeq{ \eqno(\the\eqnumber)$$ }
\def\n{\global\advance \eqnumber by 1\eqno(\the\eqnumber)}
\def\puteqno{
\global\advance \eqnumber by 1 (\the\eqnumber)}


\def\ifundefined#1{\expandafter\ifx\csname
#1\endcsname\relax}
 \newcount\sectnumber \sectnumber=0
\def\sect#1{ \advance \sectnumber by 1 {\it \the \sectnumber. #1} }

\newcount\refno \refno=0  
\def\[#1]{
\ifundefined{#1}
\advance\refno by 1
\expandafter\edef\csname #1\endcsname{\the\refno}\fi[\csname
#1\endcsname]}
\def\refis#1{\noindent\csname #1\endcsname. }

\def\label#1{
\ifundefined{#1}
\expandafter\edef\csname #1\endcsname{\the\eqnumber}
\else\message{label #1 already in use}
\fi{}}
\def\(#1){(\csname #1\endcsname)}
\def\eqn#1{(\csname #1\endcsname)}

\baselineskip=15pt
\parskip=10pt
\magnification=1200
\def\BEGINIGNORE#1ENDIGNORE{}

\baselineskip=20pt



\def\done{ {\partial \over \partial x^1}}
\def\dtwo{ {\partial \over \partial x^2}}

\def\di{ {\partial \over \partial x^i}}
\def\dj{ {\partial \over \partial x^j}}
\def\dg{ {\partial \over \partial \ga}}
\def\drho{ {\partial \over \partial \rho}}

\def\ens{\enskip}
\def\un{\underline}

\def\sect#1{{ \it #1} }
\def\un#1{\underline{#1}}
\def\hatG{  \un{\hat {\cal G}} }
\def\Q{ {\cal Q}}
\def\M{ {\cal M}}
\def\diff{ {\rm Diff}(S^1)}
\def\diffz{ {\rm Diff}_0(S^1) }
\def\ens{\enskip}

\vskip.1in
\centerline{\bf Quantum Gravity on a Circle  and  the}
\centerline{\bf Diffeomorphism Invariance of the Schr\"odinger
Equation}
\vskip.1in
\centerline{\rm R.J. Henderson and S.G. Rajeev}
\vskip.1in
\centerline{\it Department of Physics and Astronomy}
\centerline{\it University of Rochester}
\centerline{\it Rochester, N.Y. 14627}
\centerline{\it e-mail: henderson and rajeev@urhep.pas.rochester.edu}
\vskip.4in

\baselineskip=24 true pt

\centerline{\bf Abstract}

We study a model for quantum gravity on a circle in which the notion of a
classical metric tensor is replaced by a quantum metric with an inhomogeneous
transformation law under diffeomorphisms. This transformation law corresponds
to the co--adjoint action of the Virasoro algebra, and resembles that of the
connection in Yang--Mills theory. The transformation property is motivated by
the diffeomorphism invariance of the one dimensional Schr\"odinger equation.
The
quantum distance measured by the metric corresponds
to the phase of a quantum mechanical wavefunction. The dynamics of the quantum
gravity theory are specified by postulating a Riemann  metric on the space $Q$
of quantum metrics  and taking the kinetic energy operator to be the
resulting laplacian on the configuration space $Q/\rm Diff_0(S^1)$. The
resulting metric on the configuration space is analyzed and found to have
singularities. The second--quantized Schr\"odinger equation is derived, some
exact solutions are found, and a generic wavefunction behavior near one of the
metric singularities is described. Finally some further directions are
indicated, including an analogue  of  the Yamabe problem of
differential geometry.

\vfill\eject

\sect{\bf 1. Introduction}

Understanding gauge theories  such as  Yang--Mills theory and
General Relativity at the quantum level is a problem of
fundamental significance in theoretical physics.
It is useful to study these problems in simpler lower dimensional
contexts \[wittena], \[hajicek], \[rajeeva].  Yang--Mills theory on a
cylinder has been solved by
canonical methods in Ref.\[rajeev]. It would be interesting also
to study the canonical quantization of lower dimensional theories of gravity,
in analogy with that of Yang--Mills theory. The
analogy between
classical Yang--Mills theory and General Relativity has been very fruitful for
both fields. See e.g., \[dewitt]. However, this analogy is  imperfect because
the dynamical variable ( metric)  of General Relativity is a tensor while that
of Yang--Mills theory ( connection ) has an inhomogeneous transformation law.
We
will study a canonical theory of gravity in one--dimension, in which the
dynamical variable of gravity  has an inhomogeneous transformation law,
analogous to that of the connection. Since our theory is only invariant under
the diffeomorphism group of the circle
( not of the two dimensional space--time) it is to be viewed as a partially
gauge fixed theory. This is analogous to the canonical formalism of Yang--Mills
theory in the temporal gauge $A_0=0$.

We will now argue that an  inhomogeneous transformation law for the metric is
natural, if we regard the devices that measure distances to be quantum
mechanical. In a purely classical theory, the measurement of length and time
can be accomplished using rods and clocks, leading to the picture of the metric
as a tensor\[einstein]. However, in a quantum theory of matter, it would no
longer make sense to use classical objects such as  rods to measure distance:
the position of the endpoints cannot be measured with infinite accuracy.
Some quantum device must be used instead. One could use  an interference
experiment  to measure the `phase delay' of  the wavefunction between two
points.
 It would not be possible to define the length of a curve this way, since we
cannot assign a particular path to the wave; yet, it should be possible to
define a distance between two points this way. The distance between two points
then would be an average over all possible paths that connect two points rather
than the length of the minimal path.
The classical concept of distance should then be the eikonal approximation of
this quantity. It is not yet possible to implement this idea in general, but we
will be able to do so in the case of one--dimensional geometry.

  In  one dimensional Riemannian geometry (on a circle), the metric
$ds^2=g(x)dx^2$,  is determined by a  positive function $g: S^1\to R$. The
distance  between two points is $W(x_1,x_2)=\int_{x_1}^{x_2}\surd g(x)dx$ and
is
independent of the choice of the co--ordinate $x$. This can be thought of as
the solution to the eikonal form of the geodesic  equation,
\beq
\left({\pdr W(x_1,x_2)\over \pdr x_2}\right)^2-g(x_2)=0
\eeq\label{eik}
with the boundary condition $W(x_1,x_1)=0$.
If $W$ transforms as a scalar and the metric $g$ transforms as a second rank
covariant tensor, this equation is invariant under  change of
co--ordinates.
If we identify $g(x)=E-V(x)$, $E$ being the total energy and $V$ the potential
energy, this is the Hamilton--Jacobi equation of a non-relativistic point
particle of mass $m=1/2$, and the distance function $W(x_1,x_2)$ is essentially
Hamilton's characteristic function, restricted to energy $E$. By making this
connection we establish the Jacobi (or Maupertuis) principle of classical
mechanics\[lanczos]:
trajectories of a point particle of energy $E$ moving in potential $V(x)$ are
geodesics of the conformally flat metric $ds^2=(E-V(x))dx^2$. This fact is true
in higher dimensions as well, but the identification of Hamilton's
characteristic function with the distance function must be modified. The point
is that there is a correspondence between conformally flat
Riemannian geometry and the dynamics of point particles moving in potentials in
flat space. In one dimension all metrics are in fact flat so this
correspondence can be made for all one dimensional metrics.

A natural way, then, to discover a quantum generalization of one dimensional
Riemannian geometry is to base the quantum geometry on a similar correspondence
with the dynamics of non-relativistic quantum particles in flat space, these
being described by the Schr\"odinger equation:
\beq
	{\pdr^2 \psi\over \pdr x^2}+\hat q\psi=0.
\eeq\label{Hill's}
where we should identify $\hat q(x)={1\over \hbar^2}(E-V(x))$, and we will find
it convenient for later arguments to define $q(x)$ by
$\hat q=-{12\pi\over b}q+{1\over4}$ and ${b \over 12\pi}=\hbar^2$. This
differential equation with a periodic coefficient $q$ is called the Hill's
equation in the classical literature\[whittaker].

The quantum metric, in analogy to the classical situation, is $\hat q$. This
metric defines a quantum distance which may be identified with the change of
phase of the wavefunction $\psi$. In fact if we rewrite $\psi(x=x_2)$ with
boundary condition $\psi(x_1)=1$ as $\psi(x_1,x_2)=e^{{i\over
\hbar}W(x_1,x_2)}$, we find that in the classical limit, $\hbar \to 0$,
\(Hill's) reproduces the classical distance equation \(eik) with the
appropriate boundary condition. There is some
ambiguity in the choice of $\psi$ owing to the fact that the quantum distance
formula is second order, and the classical distance equation is first order. We
are free to choose a second boundary condition on $\psi$ without affecting the
classical limit. We will see later that a convenient choice is to take
$\psi(x_1,x_1+2\pi)=e^{{i\over \hbar}\rm arccos(\half \rm tr M_q)}$, where
$M_q$ is the monodromy matrix of the Hill's operator.

As mentioned, the classical Hamilton-Jacobi equation is invariant under
diffeomorphisms of the circle. To interpret the Schr\"odinger equation
\(Hill's) as a quantum distance formula it too must exhibit this symmetry.
This equation  appears, at first, not to be invariant under a change
of co--ordinates. However, if $\psi$ transforms as a half--density
\beq	\psi\mapsto \phi\circ \psi,\quad
\phi\circ\psi(x)=\psi(\phi(x))[\phi'(x)]^{-\half}
\eeq
and $q$ transforms as follows,
\beq
      q\mapsto\phi\circ q,\quad \phi\circ q(x)=q(\phi(x))\phi'^2(x)-{b\over
2\pi}\sigma_\phi(x)
\eeq
the equation is invariant under diffeomorphisms\[segal].
Here,
\beq
     \sigma_\phi(x)={1\over 6}\left[{1\over 2}{\phi'''\over \phi'}-{3\over
4}\left({\phi''\over
\phi'}\right)^2\right]+{1\over 24}(\phi'^2(x)-1).
\eeq
Thus there should be a generalization of Riemannian geometry where the metric
is not a tensor, but rather has this inhomogeneous transformation law. In
addition, the quantum generalization of the distance will no longer be a scalar
function. Taking the distance to be the phase, $W(x)$, of a solution to
\(Hill's) leads to the transformation law:
\beq
W\mapsto\phi\circ W,\quad \phi\circ W(x)=W(\phi(x))-{\hbar \over 2i}ln\phi'(x)
.\eeq
As expected, in the limit $\hbar \to 0$ ($b \to 0$), $q$ obtains the tensor
transformation property, and $W$ becomes a scalar.

The transformation law of $q$ can also be understood as  the
co--adjoint action of
the Virasoro algebra \[segal], just as the transformation law of the connection
in
Yang--Mills theory can be understood as the co--adjoint action of the affine
Kac--Moody algebra.

The configuration space of our theory of gravity is  $\Q=\{q:S^1\to R\}$, the
space of
periodic potentials or Hill's operators. We then define a Riemannian metric on
$\Q$,
\beq
	||\delta q||^2=\int \delta q^2(x) u_2^6(x) dx.
\eeq
Here $u_2$ is the solution to Hill's equation with boundary conditions,
\beq
	u_2(0)=1,\quad u_2'(0)=0.
\eeq
This metric on $\Q$ is invariant under the subgroup $\diffz$ of
diffeomorphisms that agree with the identity at $x=0$  up to third order. (
This
restriction arises for reasons which will be explained later). This metric
induces a Riemannian metric $g$ on the three dimensional manifold
$\M=\Q/\diffz$. The wavefunctions of our model for gravity on the circle  are
postulated to be functions on $\Q$ invariant under $\diffz$: i.e., functions on
$\M$. The hamiltonian the theory is the Laplace operator on $\M$ with respect
to the induced metric $g$. These choices are justified by analogy with  the
canonical formalism of Yang--Mills theory on a circle \[rajeev].

We will compute the metric tensor on $\Q$ and its  Laplace operator explicitly
in an appropriate co--ordinate system. Also, we will show that the metric $g$
has two Killing vectors and one  conformal Killing vector with constant scale
factor. This implies \[eisenhart] that $\M$   has a foliation as a
one--parameter family of two dimensional manifolds of constant negative
curvature. We will find a co--ordinate sytem that exploits this symmetry. Also
we  will show that the metric $g$ is in fact singular.
The behaviour of the eigenfunctions of the Laplace operator near the
singularities will be  determined.

\sect{\bf 2. Analogy with Yang--Mills Theory on a Circle}

In this section we follow the ideas of Segal\[segal] to construct the
 gravitational analogue of Yang--Mills theory on a circle.
 In the canonical formalism of Yang--Mills theory on a
circle,\[rajeev] the dynamical variable  is a Lie--
Algebra valued 1--form ( connection or gauge field) $A$. The invariance
group  ( gauge group) ${\cal G}$ is the loop group, the set of
functions from the circle to some compact Lie  group $G$.
Under the action of the gauge group, the
 gauge field transforms as follows:
\beq
     A\mapsto gAg^{-1}+{1\over e}gdg^{-1}.
\eeq
Here $e$ is the  gauge coupling constant. ( It  is possible to
set   $e=1$ by
rescaling $A$, but we will find it convenient not to do this.)
Infinitesimally, \beq
     \delta_\lambda A={1\over e}d\lambda +[A,\lambda]
\eeq
where $\lambda:S^1\to \un{G}$ is an element of the loop
algebra ${\cal
\un{G}}$.
It is possible to understand this as the co-adjoint
action  of the central  extension $ \un{\hat {\cal G}}$ of the
loop algebra.
\[segal].

The central extension, $\un{\hat {\cal G}}$, of the loop algebra consists of
ordered pairs,
$(\lambda,a)$, $\lambda\in {\cal \un{G}}$ and $a\in R$. The Lie
bracket of
$\un{\hat {\cal G}}$ is
\beq
     [(\lambda,a),(\tilde \lambda,\tilde a)]=
 ([\lambda,\tilde \lambda],{1\over 2\pi}\int \tr
d\lambda\tilde \lambda).
\eeq
Here $ \tr$  denotes an invariant inner product  in
$\un{G}$.
Note that elements of the form $(0,a)$ are central; i.e.,
commute with every
element. If we introduce a co--ordinate $0\leq x\leq 2\pi$
on $S^1$  and a
basis $J_{ma}(x)=(e^{imx}e_a, 0), k=(0,1)$, (where $e_a$ is a
basis in the
finite dimensional algebra $\un{G}$) we can rewrite these
commutation relations
as:
\beq
     [J_{ma},
J_{nb}]=f_{ab}^cJ_{m+n\;c}+ikg_{ab}m\delta(m+n),\quad
[J_{ma},k]=0.
\eeq
Here $g_{ab}=\tr e_ae_b$ is the inner product in the basis $e_a$.
Thus $ \hatG$ is an affine Kac--Moody
algebra\[god]. We will find
it more convenient to think of it in terms of ordered pairs as
above. The next step is to construct the  representation of $\un{\hat {\cal
G}}$ on its dual space.

But first, we digress for a technical comment on the definition of the
dual vector space
of $\hatG$. In finite dimensions, the dual of a vector space
isomorphic to
itself, although there is no natural choice of such an
identification. Each
such  choice is equivalent to the choice  of an inner product.
In infinite
dimensions, ( as for $\hatG$)in general, the dual of a vector
space is not
isomorphic to itself. If for example  we assign to  ${\cal
\un{G}}$ the topology of
 the  space  of smooth functions, its dual would be the space of
distribuitions.
This is too large for our purposes. It turns out that the best
choice \[segal]
is to work with the `smooth dual', which consists of ordered
pairs $(A,b)$, $A$
being a Lie--algebra valued 1--form on the circle and $b\in
R$. Each $(A,b)$
can be thought of as defining the linear function
\beq
     <(A,b),(\tilde \lambda,\tilde a)>=\int \tr A(x) \tilde \lambda(x) dx+
     \tilde ab
\eeq
on $\hatG$. The co--adjoint action of $(\lambda,a)$  on $(A,b)$ is
defined by the
requirement that this linear function be unchanged when we
change also $(\tilde \lambda,\tilde a)$ by the adjoint action of $(\lambda,a)$.
\beq
     <\delta_{\lambda,a}(A,b),(\tilde \lambda,\tilde a)>+
               <(A,b),[(\lambda,a),(\tilde \lambda,\tilde
a)]>=0.
\eeq
That is,
\beq
     <\delta_{\lambda,a}(A,b),(\tilde \lambda,\tilde a)>=
     -\int \tr A(x)[\lambda(x), \tilde \lambda(x)] dx-{b\over
2\pi}\int \tr d\lambda
\tilde \lambda
\eeq
which leads to
\beq
     \delta_{\lambda,a}(A,b)=(-{b\over 2\pi}d\lambda-[A,\lambda],0).
\eeq
Thus we find that $b$ is unchanged under the action of the
algebra; it can be
treated as a constant parameter. If we identify  $b={2\pi\over
e}$  we get the
familiar transformation property of a connection under a gauge
transformation.

This point of view on the gauge transformation can be
generalized to another
familiar algebra with a central extension, the Virasoro
algebra. The  gauge
theory we construct this way can be thought of as a theory of
gravity, since
the gauge group is the group of diffeomorphisms. The analogue
of the loop
algebra  $\un{\cal G}$ is the Lie algebra ${\rm Vect}(S^1)$ of
vector fields on a
circle,
\beq
     [u,\tilde u]=u{d\tilde u\over dx}-\tilde u{du\over dx}.
\eeq
Its central extension consists of ordered pairs $(u,b)$, with
$b\in  R$. The
commutation relations are
\beq
     [(u,b), (\tilde u,\tilde b)]=([u,\tilde u],
{1\over 24\pi}\int \tilde u[ {d^3u\over dx^3}+{du\over dx}]
dx).
\eeq
 The factor of ${1\over 24\pi}$ is put in to agree with the
conventional
normalization of the central extension \[god]. If we
introduce the basis $L_m=(-ie^{imx},0)$, $c=(0,1)$, we get the
familiar
commutation relations
\beq
[L_m,L_n]=(n-m)L_{m+n}+i{c\over 12}(m^3-m)\delta(m+n),\quad
[L_m,c]=0
\eeq
of the Virasoro algebra.

{}From  the co--adjoint action of the Virasoro algebra \[bowicka], \[bowickb],
\[witten], we can
find the
analogue of the gauge transformation for our approach to gravity.
Again, think of the
dual of the Virasoro  algebra  as consisting of ordered pairs
$(q,b)$, where
$q$ is a real function on the circle and $b\in R$. The pairing
is
\beq
     <(q,b),(u,a)>=\int q(x)u(x) dx +ab.
\eeq
The co--adjoint action of $(u,a)$ on $(q,b)$ is defined by
\beq
     <\delta_{u,a}(q,b),(\tilde u,\tilde a)>+
               <(q,b),[(u,a),(\tilde u,\tilde a)]>=0.
\eeq
This means that
\beq
     <\delta_{u,a}(q,b),(\tilde u,\tilde a)>=
     -\int q(x)(u\tilde u'-\tilde u u') dx-
{b\over 24\pi}\int  \tilde u(u'''+u') dx
\eeq
which leads to
\beq
     \delta_{u,a}(q,b)=(-{b\over 24\pi}(u'''+u')+2qu'+q'u,0).
\eeq
Again, note that $a$ has no effect on $q$ and that $b$ remains
unchanged.

 To see the geometrical meaning of this transformation, it is
useful to
consider the limit $b=0$. Then the inhomogeneous term
disappears, and we find
that $q$ transforms as a tensor of order two. That is,
$q(x)dx^2$ is invariant
under diffeomorphisms  when $b=0$. Thus $q$ can be thought of
as a  metric
tensor on $S^1$.   In this case, all metrics are of course
flat. The
diffeomorphism invariant information contained in the metric
is the length
$\int \surd q(x) dx$ of the circle.

We will be mostly interested in the case $b\neq 0$. Thus our
$q$ has an
inhomogenous transformation property,
\beq
     \delta_{(u,a)}q=-{b\over 24\pi}(u'''+u')+2qu'+q'u.
\eeq
Our theory can therefore be thought of as  a generalization of
one--dimensional
Riemannian geometry: the metric is not a tensor, but rather
has an affine
transformation law analogous to that of a connection. There should be
analogues of this
transformation in higher dimensions, but they are as yet too
difficult to find.

Segal\[segal]   has found the transformation law of $q$
under finite
diffeomorphisms as well. Remarkably,  it involves the
Schwarzian derivative
which plays a central role in classical analysis. (See
Ref.\[bieber] for a geometric interpretation of the Schwarzian derivative)

If $\phi:S^1\to S^1$  is a diffeomorphism, $q$ transforms
under it to $\phi\circ q(x)$ where,
\beq
      \phi\circ q(x)=q(\phi(x))\phi'^2(x)-{b\over 2\pi}\sigma_\phi(x).
\eeq\label{phiq}
Here,
\beq
     \sigma_\phi(x)={1\over 6} S_\phi(x)+{1\over
24}(\phi'^2(x)-1)
\eeq
and $S_\phi$ is the Schwarzian derivative of $\phi$:
\beq
     S_\phi=\half {\phi'''\over \phi'}-{3\over
4}\left({\phi''\over
\phi'}\right)^2.
\eeq
The quantity $\sigma_\phi$ satisfies the identity
\beq
\sigma_{\phi\circ\psi}(x)=\sigma_{\phi}(\psi(x))\psi'^2(x)+
\sigma_\psi(x)
\eeq
which is necessary in order that the action of the
diffeomorphism on $q$
satisfy the composition law.

The term ${1\over 24}(\phi'^2-1)$ in the transformation law can be removed
by adding a constant, -${b\over 48\pi}$ to $q$. This  amounts to a different
choice of central term in the Virasoro algebra. However, our choice is more
convenient,  since then the isotropy group of $q=0$ is the group of Mobius
transformations. Under infinitesimal diffeomorphisms, the condition to leave
$q=0$ invariant is $ u'''+u'=0$, which has solutions
$u=(a-\bar a)L_0+bL_1-\bar b L_{-1}$, where $L_m=-ie^{imx}$. This generates
the subgroup of Mobius
transformations on which the central term of the Virasoro algebra vanishes.
The finite transformations  which leave $q=0$ invariant are of the form
$x\mapsto \phi(x)$,
\beq
	e^{i\phi(x)}={Ae^{ix}+B\over \bar B e^{ix}+\bar A},\quad
    A\bar A-B\bar B=1.
\eeq
This subgroup of diffeomorphisms is often also  called $SU(1,1)$ or $PSL_2(R)$.
If we had used a
different choice of  the constant in $q$ ( hence of
$\sigma_\phi$), the point in $\Q$ with $PSL_2(R)$ as the isotropy group would
not have been  the origin.

In the case of gauge fields the subgroup that preserves the
trivial connection
$A=0$ consists of global or constant gauge transformations. In
our case, from
the above observation,  the analogous subgroup is
$PSL_2(R)$.

In the case of gauge theory, we are interested in the space of
connections
${\cal A}$ modulo gauge transformations. However, only the
subgroup ${\cal
G}_0$ of  gauge transformations that are equal to the identity
at $x=0$ acts
without fixed points. The space ${\cal A}/{\cal G}_0$ is a
manifold and the
wavefunctions of pure gauge theory can be  viewed as functions
on this space.
On a circle, this quotient space is  finite dimensional: the
only gauge
invariant observable is the parallel transport operator (
Wilson loop) around
the circle.
This quantity can  be defined precisely using  the
differential
equation ( parallel transport equation)
\beq
     {d\psi\over dx}+eA(x)\psi(x)=0.
\eeq
This equation  is invariant under the transformations
$\psi(x)\mapsto \rho(g(x))\psi(x)$ (where $\rho$ is a
representation of $G$),and
$A\mapsto gAg^{-1}+{1\over e} gdg^{-1}$. It is enough to
consider the fundamental representation, since solutions in
the general case can be obtained by taking products of the
fundamental solution.
Although $A(x)$ is periodic, the
solution is not, in general, periodic. The solution instead
satisfies the condition
\beq
     \psi(2\pi)=\rho(W(A))\psi(0)
\eeq
where $W:{\cal A}\to G$ is the `parallel transport operator'
or Wilson loop. Given $W(A)$,  $A$  is determined up to an
action of the gauge group ${\cal G}_0$. In fact we can
identify the quotient space ${\cal A}/{\cal G}_0$ with the
group $G$.  It is now possible to show that ${\cal A}$ is the
total space of   a principal fibre bundle
\beq
     {\cal G}_0\to {\cal A}\to G
\eeq
with $G$ as the base space.  Wavefunctions of Yang--Mills
theory are functions on ${\cal A}$ invariant under the action
of ${\cal G}_0$, hence are just functions on $G$.

An analogue of this construction can be found following
Segal's ideas. The analogue of ${\cal A}$ is ${\cal
Q}=\{q:S^1\to R\}$. The gauge  group  ${\cal G}$ is replaced
by the group $Diff(S^1)$ of diffeomorphisms of the circle. It
acts on $\Q$ by the above transformation.
The analogue $Diff_0(S^1)$ of ${\cal G}_0$ is given by
diffeomorphisms of $S^1$ satisfying
\beq
     \phi(0)=0,\quad \phi'(0)=1,\quad \phi''(0)=0.
\eeq
That is, they agree with the identity up to three derivatives.
At least for infinitesimal transformations in $Diff_0(S^1)$,
it is easy to see that there are no fixed points in $\Q$. If
the point $q\in \Q$ is invariant under the action of a vector
field $u$,
\beq
 q'u+2qu'-{b\over 24\pi}(u'''+u')=0.
\eeq
This third order equation has to be solved with the boundary
conditions $u(0)=u'(0)=u''(0)=0$. Clearly $u(x)=0$ is the only
solution. The proof in the case of finite transformations
looks more complicated; we are not able to provide one. We
expect the quotient space $\Q/Diff_0(S^1)$ to be well--
defined.

Now we look for the analogue of the parallel transport
operator $W(A)$.  Another observation of Segal is that
the equation
\beq
D^{(2)}_{\hat q}\psi=\psi''+\hat q \psi =
\psi''+(-{12\pi\over b}q+{1\over 4})\psi=0
\eeq
is invariant if $q$ transforms as above and $\psi$ transforms
as a  density of weight $-\half$. $\psi$ being a half density means that
$\psi(x) (dx)^{-
\half}$ is invariant. Equivalently, $\psi\to \phi\circ \psi$,
where,
\beq
     \phi\circ\psi(x)=\psi(\phi(x))[\phi'(x)]^{-\half}.
\eeq
More generally, there is a differential operator, $D^{(2s+1)}_{\hat q}$, of
order $2s+1$ which maps densities of weight $-s$ to those of weight
$s+1$, for $s=0,1/2,1,3/2,...$\[scherer]. These operators are analogous  to the
covariant derivatives of gauge theory. We note in
particular that the change of $q$ under an infinitesimal
diffeomorphism, $\phi (x)=x+u(x)$, is proportional to:
\beq
     D^{(3)}_{\hat q} u=   u'''+u'-{24\pi \over b}(2qu'+q'u).
\eeq
where $D^{(3)}_{\hat q}$ is the differential operator which maps vector fields
(weight = -1) to quadratic forms (weight = 2).

The solutions of the homogeneous equations $D^{(2s+1)}_{\hat q}u=0$ are just
the products of the $2s$ solutions of $D^{(2s)}_{\hat q}u=0$.
Thus, the densities of weight $-\half$
( which can be thought of as spinors) form  the analogues of
the fundamental representation of $\diffz$.

The solutions of the equation ( known as  Hill's equation)
\beq
     \psi''+\hat q\psi=0
\eeq
( with $\hat q=-{12\pi\over b}q+{1\over 4}$)
are not in general  periodic, even though $q$ itself is
periodic.  In the special case $q=0$, there are anti--periodic solutions to
Hill's equation.
In general,  a basis $u_1,u_2$ of  solutions, will
change by  a linear transformation $M_q$ as we go from $x=0$
to $x=2\pi$:
\beq
     \pmatrix{u_1(x+2\pi)\cr u_2(x+2\pi)}=M_q
\pmatrix{u_1(x)\cr u_2(x)}.
\eeq
A standard basis  is  defined by the boundary conditions,
\beq
	u_1(0)=0, u_1'(0)=1;\quad u_2(0)=1, u_2'(0)=0.
\eeq
In this basis,  matrix
\beq
M_q=\pmatrix{u_1'(2\pi)&u_1(2\pi)\cr
	     u_2'(2\pi)& u_2(2\pi)}
\eeq
is invariant under $\diffz$,
since this subgroup preserves the boundary conditions.
 $M_q$ is of determinant one, since the Wronskian of
the two solutions is one. This is the analogue of the
parallel transport operator $W(A)$. ( As Segal points out, we can get
slightly more information from the differential equation: an
element of the universal covering group $\tilde SL_2(R)$,
rather than one of $SL_2(R)$. Our considerations will not
distinguish between these.)

The conjugacy class of $M_q$ in $SL_2(R)$  is invariant under the full
diffeomorphism group, not just the subgroup $\diffz$.
This conjugacy class is labelled by the trace of $M_q$. If
this trace is greater than two, $M_q$ is said to be
hyperbolic. Then, $M_q$ has real eigenvalues and the
solutions to Hill's equation, when extended to the real line, will blow up at
either $x=\infty$ or $x=-\infty$. If the trace is equal to
two, $M_q$ cannot be diagonalized  unless $M_q=1$. There will
be in this case an eigenvalue equal to one, and there will be a periodic
solution to Hill's equation. If $M_q=1$, both solutions are periodic. If $\tr
M_q<2$, $M_q$ is said to be elliptic and the eigenvalues are
complex numbers of modulus one. Then the solutions to Hill's equation will
remain finite when extended over the real line: they are
similar to the Bloch waves in a solid.

As discussed above, we can regard Hill's equation as the Schr\"odinger
equation of a particle in a periodic potential, in which case we identify
${b\over 12\pi}\hat q(x)=E-V(x)$, $E$ being the total energy and $V(x)$ is the
potential energy. Then the parameter ${b\over 12\pi}$ is analogous to
$\hbar^2$. In the limit $b\to 0$,
we can use the WKB ( eikonal) approximation to find $\tr
M_q=2\cos(\int_0^{2\pi} \surd \hat q(x)dx)$. Thus the invariant $\tr
M_q$ is, in this limit, determined by the length of the circle
with respect to the Riemannian metric $\hat q(x)dx^2$. We have therefore a
direct
correspondence between the diffeomorphism invariant quantities of the classical
and quantum theories of the geometry of a circle.

Since the wavefunction $\psi$ transforms as a density, its
change in phase over the length of the circle is diffeomorphism invariant, and
is therefore a function of $\tr M_q$. We can in fact make a choice of boundary
conditions such that the total phase change over the
length of the circle is
$\arccos[\half \tr M_q]$. Then the quantum length of the circle reduces to the
classical length in the limit $\hbar \to 0$. This choice of boundary conditions
resolves the ambiguity in $\psi$ discussed earlier. For example, the quantum
distance of the point $x$ from the origin is the phase of $\psi(x)=\psi(0,x)$,
where $\psi$ satisfies Hill's equation with the boundary conditions $\psi(0)=1$
and $\psi(2\pi)=e^{{i\over \hbar}\arccos(\half \tr M_q)}$.

The phase of the wavefunction $\psi$ will be real in the classically allowed
regions ($E-V(x)>0$), so it is in these regions that the quantum measure of
distance will be real. In the classically forbidden regions ($E-V(x)<0$) the
quantum distance will take on imaginary values. Finally in those regions where
$E-V(x)=0$, the metric $\hat q(x)$ disappears, and all distances are null.

We find then that in the quantum theory the potential (or more precisely $\hat
q(x)={1\over \hbar^2}(E-V(x))$), no longer transforms as a tensor, but still
appears to have a geometrical meaning. It would be interesting to find higher
dimensional analogues of this quantum generalization of the metric tensor.

The quantity
\beq
	F_q(x)={u_1(x)\over u_2(x)}.
\eeq
is a local diffeomorphism of $S^1$ to $RP^1$.
It is not in general a diffeomorphism, since $F_q(2\pi)\neq F_q(0)$.
In fact, $F_q(2\pi)={M_{12}\over M_{22}}$  is an invariant and under $\diffz$.
Also, $F_q'(x)={1\over u_2^2(x)}$ transforms under $\diffz$ as a density of
weight
one. These are  useful quantities in what follows.

\sect{\bf 3. The Hamiltonian}

So far we have described the kinematical aspects of the
theory, obtained by analogy with Yang--Mills theory. Now we
will go beyond this and look for a dynamical principle, again
by analogy with Yang--Mills theory.

The space ${\cal A}$ has a   metric
\beq
     ||\delta A ||^2=\int \tr \delta A(x) \delta A(x) dx
\eeq
invariant under  the action of the gauge group. In this case,
this metric is flat. The true configuration space  is the
quotient $G={\cal G}/{\cal G}_0$, to  which the above metric
can be projeted, since it is gauge invariant. The metric so
obtained on $G$ is the usual Cartan--Killing metric,
corresponding to the invariant bilinear $\tr$ on the Lie
algebra $\un{G}$.

 This leads to a natural classical dynamics, where  the
classical trajectories are geodesics with respect to this
metric. In the case of Yang--Mills theory on a cylinder, this
is precisely the dynamics determined by the Yang--Mills
equations of motion.  Of course, Yang--Mills equations do not
in general describe geodesics.  In 1+1 dimensions, the
potential energy term in the hamiltonian is absent,  so that
the classical trajectories are geodesics.
This can be most easily seen by considering the Yang--Mills
action in first order form
\beq
     S=\int \tr E(x){\pdr A(x)\over \pdr t}dx dt+
               \int \tr A_0\{dE+[A,E]\} dxdt
                    -\half  \int \tr E(x)^2 dxdt.
\eeq
The first term identifies $E$ and $A$ as canonical conjugates
of each other. The second term imposes the first class
constraint
\beq
     dE+[A,E]=0
\eeq
which generates the gauge transformations. The last term is
the hamiltonian.

In  quantum Yang--Mills theory, the wavefunctions  are
functions on  the true configuration space $G$. The
hamiltonian is just the Laplace operator on $G$. The resulting
quantum dynamics can be easily solved using group theory\[rajeev].

To find an analogous theory invariant under the diffeomorphism
group, we must look for a metric on $\Q$. In the limit $b=0$,
the simplest choice is \[polyakov]
\beq
     ||\delta q||_0^2=\int \delta q^2(x) q^{-{3\over 2}} dx.
\eeq
This already is not  a flat metric on $\Q$. For general value of $b$, we
look for a metric of the form
\beq
     ||\delta q||^2=\int \delta q^2(x) G_q(x) dx
\eeq
where $G_q$ has to transform as a density of weight $-3$ as
$q$ transforms as  in \(phiq). We can build such a quantity
from a solution to Hill's equation, which transforms as a
density of weight $-\half$. However, the solution is not
unique; a choice of boundary condition is necessary. We are
not able to find a choice that is invariant under the full
diffeomorphism group. If we pick the solution $u_2$ with the
boundary condition
\beq
     u_2(0)=1,\quad u_2'(0)=0
\eeq
the resulting metric
\beq
     ||\delta q||^2=\int \delta q^2(x) u_2^{6}(x) dx.
\eeq\label{metric}
will be invariant under $\diffz$. This is because the
boundary conditions on $u_2$ remain invariant under
$\diffz$.
This means that the point $x=0$ has a special significance in
our theory. (This might be natural if we regard $S^1$ as
obtained by compactifying the real line, the point $x=0$ then
corresponds to infinity on the real line.) There could also be
more complicated metrics that are invariant under under the
full diffeomorphism group, but we believe a first
investigation should focus on the simplest possibility.
Even this choice is not a flat metric on $\Q$. It  reduces to
$||\delta q||_0^2$ (up to a constant scaling factor)
in the limit $b\to 0$, as can be seen using
the WKB approximation.
Since $F_q'={1\over u_2^2}$, we can also write the metric as
\beq
	||\delta q||^2=\int \delta q^2(x) F_q'^{-3}(x)dx
\eeq

Due to the  invariance under $Diff_0(S^1)$, we have an induced
 metric on the quotient space ${\cal M}=\Q/Diff_0(S^1)$. By
analogy to Yang--Mills theory ,we postulate that the
wavefunctions of our theory are  functions on $\Q$ invariant
under $Diff_0(S^1)$; i.e., functions on the quotient space
${\cal M}$. The hamiltonian operator is the Laplacian on
${\cal M}$ with respect to the induced Riemannian metric. In
the classical limit, the trajectories will be the geodesics of
this metric. We do not yet know whether this dynamics can be
derived from an action principle analogous to the Yang--Mills
action.

We will now study the metric on ${\cal M}$ and the
corresponding dynamics more explicitly.

\sect{\bf 4. The  Riemannian metric on ${\cal M}$}

It is useful to think of the quotient  ${\cal M}$ as  the base
space of a Principal Fibre Bundle $\diffz\to \Q\to \M$.
although it is difficult to give rigorous definitions and
proofs, this geometric language is very suggestive as in the
case of Yang--Mills theory. A vertical vector at   $q\in \Q$
represents an infinitesimal gauge transformation
\beq
     \delta_u q(x)=-{b\over 24\pi}D_{\hat q}^{(3)}u=
     -{b\over 24\pi}(u'''+u')+2qu'+q'u.
\eeq
A 1--form, $\xi_q$, at $q\in \Q$ may be said to be horizontal
if it annihilates all the
vertical vectors at $q$. This leads to the differential equation,
\beq
	D_{\hat q}^{(3)}\xi_q=0.
\eeq\label{hor}
If $q=0$, the solutions are straightforward:
\beq
	\xi_0(x)=a_1\sin x +a_2\cos x +a_3.
\eeq
More generally, we note that the solutions to the equation \(hor) are given by
products of solutions to Hill's equation. (This can be verified by
straightforward calculation, but can also be understood in terms of the
construction of the sequence of operators $D_q^{(n)}$ in Ref. \[scherer].)
Hence
\beq
	\xi_q(x)=au_2^2(x)+bu_1(x)u_2(x)+cu_1^2(x)=[a+bF_q+cF_q^2]F_q'^{-1}
\eeq
 $ a,b,c$ being constants.

We will now show that, a {horizontal vector} $q\in \Q$ can be defined to be one
that is orthogonal
to all the vertical vectors. This defines a connection in the principal bundle
$\diffz\to \Q\to \M$. In fact a horizontal vector can be obtained by operating
on a horizontal 1--form by the metric \(metric):
\beq
	\delta_h q(x)=F_q'^3\xi_q =[a+bF_q(x)+cF_q^2(x)]F_q'^2(x).
\eeq
The length of such a horizontal vector can now be computed to be
\beq
	||\delta_h q||^2=\int_0^{2\pi} \delta_h
q^2(x)F_q'^{-3}(x)dx=\int_0^{\gamma}[a+bx+cx^2]^2dx
\eeq
where
\beq
	\gamma=F_q(2\pi)={u_1(2\pi)\over u_2(2\pi)}={M_{12}\over M_{22}}.
\eeq

A horizontal vector can be thought of as tangential to the base manifold $\M$
of our Principal bundle. A point in $\M$ is parametrized by a matrix $M_q\in
SL_2(R)$. A vector in   $\M$  is an infinitesimal variation
\beq
	\delta M_q=\pmatrix{\de u_1'(2\pi)&\de u_1(2\pi)\cr \de u_2'(2\pi)&\de
u_2(2\pi)}
.\eeq\label{delM matrix}\noindent
The corresponding horizontal vector in $\Q$ ( the ` horizontal lift') can be
found using the first order perturbation of the Hill's equation:
\beq
D_{\hat q}^{(2)}\de u_i=-\de_h\hat qu_i\qquad\quad i=1,2
\eeq\label{pertHill}\noindent
The boundary conditions are $\delta u_i(0)=\delta u_i'(0)=0$.
This equation can be solved using the Green's function of Hill's operator:
\beq
G_{\hat q}(x,y)=\cases{u_1(y)u_2(x),&$x \le y$\cr
u_1(x)u_2(y),&$x \ge y$\cr}
\eeq
so that
\beq
 \quad \de u_i(x)=-{\int_0^{2\pi}dy \ens
G_{\hat q}(x,y)\de_h q(y )
u_i(y)}+A_iu_1(x)+B_iu_2(x)\qquad i=1,2
\eeq
where $A_i$ and $B_i$ are determined by the boundary conditions on $\de u_i$.
We will get
\beq
	A_i=0\quad B_i=\int_0^{2\pi} dx u_1(x) \delta_h q(x) u_i(x),\quad i=1,2.
\eeq
Thus $\delta_h q$ at $q\in \Q$  projects to the vector
\beq
	\delta M_q=\Delta M_q
\eeq where
\beq
	\Delta=\pmatrix{-\eps_{21}&\eps_{11}\cr
-\eps_{22}&\eps_{12}}
\eeq
\beq
\hbox{with}\qquad \eps_{ij}=\eps_{ji}={\int_0^{2\pi}dx \ens
u_i(x)\de_hq(x)u_j(x)}\qquad i,j=1,2
.\eeq
Using the earlier expressions for $\delta_h q$ and for $F_q$ in terms of $u_1$
and $u_2$, we get
\beq
	\Delta=a\Ga_1+b\Ga_2+c\Ga_3
\eeq\label{del gam}\noindent
where the 2x2 matrices $\Ga_i$ can be written in the following way:
\beq
\pmatrix{\Ga_1\cr \Ga_2\cr \Ga_3\cr}=
\pmatrix{\ga &\ga^2/2&\ga^3/3\cr \ga^2/2&\ga^3/3&\ga^4/4\cr
\ga^3/3&\ga^4/4&\ga^5/5\cr}
\pmatrix{\tau_1\cr \tau_2\cr \tau_3}.
\eeq\label{ga vs tau}\noindent
Also, $\{ \tau_i\}$ gives a basis for $\un{SL_2(R)}$:
\beq
\tau_1=\pmatrix{0&0\cr -1&0},\quad
\tau_2=\pmatrix{-1&0\cr 0&1},\quad
\tau_3=\pmatrix{0&1\cr 0&0}
\eeq

Now we can find explicitly the metric tensor on $\M$ obtained by projecting
from $\Q$. It is useful to identify $\M$ with $SL_2(R)$ locally, and introduce
the usual Maurer--Cartan forms,
\beq
\delta M_q M_q^{-1}=\omega^1\tau_1+\omega^2\tau_2+\omega^3\tau_3.
\eeq
We can now  find the length of $\delta M$ by expressing it in terms of
$\delta_h q$ and using \(metric):
\beq
	||\delta M_q||^2=G_{ij} \omega^i\omega^j.
\eeq
Here $G$ is the $3\times 3$ matrix, whose inverse is given by
\beq
G^{-1}=\pmatrix{\ga &\ga^2/2&\ga^3/3\cr \ga^2/2&\ga^3/3&\ga^4/4\cr
\ga^3/3&\ga^4/4&\ga^5/5\cr}
\eeq
so that
\beq
	G=3\pmatrix{3/\ga &-12/\ga^2&10/\ga^3\cr -12/\ga^2&64/\ga^3&-60/\ga^4\cr
10/\ga^3&-60/\ga^4&60/\ga^5\cr}.
\eeq\label{G}\noindent
Thus the metric tensor  on $\M$ induced by \(metric) is
 \beq
g=G_{ij}\omega^i\otimes \omega^j.
\eeq\label{g}\noindent

\sect{\bf 5. Symmetries of the Metric on $\M$}

Our model quantum gravity on a circle has the Laplacian of the metric on $\M$
as the hamiltonian. Clearly, to understand this better we must exploit the
symmetries of the metric. We will show next that there are two Killing vectors
and one conformal Killing vector.
Methods of classical differential geometry \[eisenhart] can now be used to find
a co--ordinate system adapted to the metric. Unfortunately we will not be able
to separate the variables in the eigenvalue equation for the Laplacian; but we
will achieve a considerable simplification. We will also discover that the
metric on $\M$ has singularities. This is  an indication that our quantum
gravity theory has divergences, which may require a renormalization.

The Maurer--Cartan forms $\omega^i$ are invariant under the right
multiplication by any  element  of $SL_2(R)$. If the coeffiecients $G_{ij}$ had
been constants, any such right action would have been an isometry of our metric
$g$.  Although $G_{ij}$ are not constant in our case, they  only depend on the
variable $\gamma$. Hence a right translation that leaves $\gamma$ invariant,
will be an isometry.
Now,
\beq
	\gamma={M_{12}\over M_{22} }
\eeq
is unchanged under right multiplication by elements of the form
\beq
	\pmatrix{\alpha&0\cr  \beta&\alpha^{-1}}.
\eeq
Thus there are two independent Killing vectors, corresponding to
infinitesimally small transformations of the above type.
Furthermore, note that the matrix elements of $G_{ij}$ are monomials in
$\gamma$ of varying degrees. This leads to a homothety ( conformal isometry
 with constant scale factor ) of the metric ( see below).

Let us define the vector fields

\beq
	W_if(M_q)=\lim_{t\to 0}{f(M_q[1-t\tau_i])-f(M_q)\over t},\qquad i=1,2,3
\eeq
Then, $W_1$ and  $W_2$ are Killing vectors. It will also be convenient to
introduce the generators of the left action ( right invariant vector fields)
\beq
	V_if(M_q)=\lim_{t\to 0}{f([1+t\tau_i]M_q)-f(M_q)\over t},\qquad i=1,2,3
\eeq
These satisfy
\beq
	[V_i,V_j]=f^k_{ij}V_k,\quad [V_i,W_j]=0,\quad [W_i,W_j]=f^k_{ij}W_k
\eeq
where $f^k_{ij}$ are the structure constants of $\un {SL_2(R)}$, using
the basis $\{\tau_i\}$. The vector  fields $V_i$ are dual to the Maurer--Cartan
forms $\omega^i$,
\beq
	i_{V_i}\omega^j=\de^j_i.
\eeq

We can parametrize $M_q$ by the co--ordinate system $(\alpha,\beta,\gamma)$,
defined by
\beq
M_q=\pmatrix{1 &\ga \cr 0 &1\cr}\pmatrix{\al &0\cr \be &\al^{-1}\cr}
\eeq\label{coord}
This is valid near the identity element of $SL_2(R)$. Our manifold $\M$ agrees
with  $SL_2(R)$ only in a local neighborhood of the identity, so this is
sufficient. We will now find a  series of co--ordinate transformations aimed at
 finding a co--ordinate system that exploits the symmetries of $g$ fully.

The Killing vectors $W_1,W_2$ satisfy
\beq
	[W_1,W_2]=-2W_1.
\eeq\label{commW}\noindent
There is then \[eisenhart] a pair of variables $(x^1,x^2)$  such that
\beq
W_1=e^{2x^2}\done \qquad,\qquad W_2=\dtwo
.\eeq
One such  co--ordinate system is given by
\beq
x^1=-\be /\al,\quad x^2=-ln\al,\quad x^3=\ga
\eeq
 for $\alpha>0$.
 We then find the right-invariant vector fields:
\beq
\eqalign{&V_1= (1-2\ga x^1)\done - \ga \dtwo +\ga^2 \dg \cr
&V_2=2x^1 \done + \dtwo - 2\ga \dg \cr
&V_3=\dg \cr}
\eeq
and the dual 1-forms:
\beq
\eqalign{&\om^1=dx^1-2x^1dx^2\cr
&\om^2=\ga dx^1+(1-2x^1\ga )dx^2\cr
&\om^3=\ga^2dx^1+2\ga(1-\ga x^1)dx^2+d\ga \cr}
\eeq
or
\beq
\eqalign{&\om^2=\ga\om^1+dx^2\quad \cr
&\om^3=d\ga + 2\ga \om^2-\ga^2\om^1\quad \cr}.
\eeq

Consider the vector field
\beq
	D=\ga{\pdr \over \pdr\ga}-x^1{\pdr \over \pdr x^1}.
\eeq
It generates a scale transformation in which $\gamma$ has degree one, $x^1$ has
degree $-1$ and $x^2$ has degree zero. Then, $\omega^1$ has degree $-1$,
$\omega^2$ has degree zero and $\omega^3$ has degree one. That is, $\omega^i$
has degree $i-2$.
 Since $G_{ij}$ is a monomial in $\gamma$  of degree $-(i+j-1)$, it follows
that
the metric tensor $g=G_{ij}\omega^i\otimes \omega^j$ is of degree $-3$. That
is,
\beq
	{\cal L}_Dg=-3g.
\eeq
Thus $D$ is a conformal Killing vector with constant scale factor ( or
infinitesimal homothety). Thus our theory admits a scale invariance. It would
be interesting to see if this scale invariance is related to the divergences
that arise in the quantum theory ( see below). Also, there could be a
connection
to the renormalization theory of quantum mechanics\[rengrp].

The surfaces $\gamma$=constant are two dimensional sub--manifolds of
constant curvature. This is because, $W_1$ and $W_2$ are tangential to these
manifolds and form two independent Killing vectors satisfying \(commW)
\[eisenhart]. However, the vector ${\pdr\over \pdr\gamma}$ is {\it not}
orthogonal
to this surface. If we find a vector $N$ that is orthogonal, we will be able to
bring the metric into block--diagonal form. It is not difficult to find a
vector that is normal to $W_1$ and $W_2$:
\beq
N=(1-{3x^1\ga \over 2})\done - {3\ga \over 4}\dtwo + {3\ga^2 \over 5}\dg
\eeq
The integral curves of this vector field are,
\beq
x^1={10\over 9\ga}+{\xi^1\over \ga^{5/2}},\quad
x^2=-{5\over 4}ln\ga -\half \xi^2.
\eeq
Here, we regard $\gamma$ as the parameter that varies along the curve. The
quantities $\xi^1,\xi^2$ are constants of integration, which distinguish
between different integral curves. If we use $\gamma, \xi^1,\xi^2$ as the
co--ordinates, the metric will be block--diagonal. Moreover, if we trade
$\gamma$ for
\beq
\rho =-{2\sqrt{5}\over 3\ga^{3/2}}
\eeq
the metric will take the form
\beq
g=d\rho^2 + A\rho^4d\xi^1d\xi^1 + 2\rho^3 hd\xi^1d\xi^2 +
{\rho^2\over A}(h^2 + q)d\xi^2d\xi^2.
\eeq
Here,
\beq
h=A\rho \xi^1 + B\qquad,\qquad q=AC-B^2.
\eeq\label{h and q defs}\noindent
And, the constants $A,B,C$ are:
\beq
A=-{93\over 200},\qquad B={27\over 5\sqrt{5}},\qquad C={43\over 5}
\eeq
We also note for future use the contravariant elements:
\beq
\eqalign{&g^{11}={1\over Aq\rho^4}(h^2 + q)\cr
&g^{12}=g^{21}=-{h\over q\rho^3}\cr
&g^{22}={A\over q\rho^2}\cr}
\eeq\label{contra elements}\noindent
and:
\beq
\tilde{g} \equiv detg_{ij}= q\rho^6
\eeq
As mentioned previously the two-dimensional submanifolds $\rho=$ constant,
which contain the two Killing vectors, are in fact surfaces of constant
negative curvature. We find that the intrinsic curvature of these surfaces is
$R_{(2)}=-A/q\rho^2$.

This is the simplest form to which  we can  bring the metric.  The Killing
vectors $W_1,W_2$ now take the form
\beq
	W_1=e^{-2\xi^2}{\pdr\over \pdr \xi^1}\quad W_2=-2{\pdr\over \pdr \xi^2}.
\eeq

Finally, \(h and q defs)-\(contra elements) enable us to write the action of
the
Laplacian on $\Psi \in C^{\infty}(M)$. It is clear that translations in $\xi^2$
are symmetries of the Laplacian, so that we can assume that the wavefunctions
satisfy
\beq
\partial \Psi/ \partial \xi^2=K_2\Psi.
\eeq
and let $\Psi(\xi^1,\xi^2,\rho)\mapsto e^{K_2\xi^2}\Psi(\xi^1,\rho)$.
Then, we can express the Laplacian as:
\beq
\eqalign{\L \Psi &= {1\over \sqrt{\tilde{g}}}\di (\sqrt{\tilde{g}}g^{ij}\dj
\Psi) \cr &={1\over \rho^3}\drho (\rho^3{\partial \Psi \over \partial \rho })
+ {h^2+q\over Aq\rho^4}{\partial^2\Psi \over (\partial \xi^1)^2}
-{2h(K_2-1)\over q\rho^3}{\partial \Psi \over \partial \xi^1}
+{AK_2(K_2-1)\over q\rho^2}\Psi \cr}
\eeq \label{L action}\noindent

This is a partial differential operator in two variables, and we are not able
to separate the variables any further. However, the form is simple enough that
we can study some of its qualitative  behaviour. We can also find some special
solutions to the resulting Schr\"odinger equation.

\sect{\bf 6. Metric Singularities and Wavefunction Behavior in Their Vicinity}

We have seen that the metric coefficients (either covariant or contravariant)
have singular behavior for $\rho \mapsto 0$ and for $\rho \xi^1 \mapsto
\infty$. We have also calculated the intrinsic curvature of the $\rho=$
constant
surfaces and found that it blows up for $\rho \mapsto 0$.
These facts alone however do not imply the presence of true metric
singularities. The singular behavior of the metric coefficients may just be due
to the choice of coordinate system. The singular behavior of the intrinsic
curvature of some submanifolds may be cancelled by a similarly singular
behavior of the submanifolds' extrinsic curvature. For this reason we have
performed the somewhat lengthy task of calculating the Ricci scalar curvature
of g explicitly, finding it to be:
\beq
R={1\over \rho^2}(R_4 h^4+R_3 h^3+R_2 h^2+R_1 h+R_0)
\eeq
where $R_i$ are constants.

Thus the metric g does indeed have singularities at $\rho =0$ and
$\rho \xi^1=\infty$. We
can investigate the effect on wavefunctions of the singularity at $\rho = 0$
by examining the Schr\"odinger equation, $\L \Psi = E\Psi$, in the
limit $\rho \mapsto 0$. Using \(L action) we find that for $\rho \xi^1\to
0$ the Schr\"odinger equation approaches:
\beq
{1\over \rho^3}\drho (\rho^3{\partial \Psi_0 \over \partial \rho })
+ {C\over q\rho^4}{\partial^2\Psi_0 \over (\partial \xi^1)^2}
-{2B(K_2-1)\over q\rho^3}{\partial \Psi_0 \over \partial \xi^1}
+{AK_2(K_2-1)\over q\rho^2}\Psi_0
=E\Psi_0
\eeq\label{psi zero}\noindent
i.e., $\Psi (\rho,\xi^1)\sim \Psi_0(\rho,\xi^1),\quad \rho \xi^1\to 0$.

This equation for $\Psi_0(\rho,\xi^1)$ is separable. We may assume that
$\partial \Psi_0/ \partial \xi^1=K_1\Psi_0$ such that \(psi zero) becomes:
\beq
\Psi_0''+{3\over \rho}\Psi_0'
+\biggl({CK_1^2\over q\rho^4}-{2BK_1(K_2-1)\over q\rho^3}+
{AK_2(K_2-1)\over q\rho^2}\biggr)\Psi_0
=E\Psi_0
\eeq\label{psi zero 2}
upon letting $\Psi_0(\rho,\xi^1) \mapsto e^{K_1\xi^1}\Psi_0(\rho)$.

Consider first the case $K_1=0$, such that $\Psi_0(\rho,\xi^1)$ is independent
of
$\xi^1$. Then \(psi zero 2) is related to Bessel's equation and has the two
independent solutions:
\beq
\Psi_0(\rho)=\cases{{1\over \rho}J_{{+\atop -}\nu }(i\sqrt{E}\rho ),&if
$\nu \notin Z$\cr
{1\over \rho}J_{\nu}(i\sqrt{E}\rho),
{1\over \rho}N_{\nu}(i\sqrt{E}\rho),&if
$\nu \in Z$\cr}
\eeq\label{bessel}
where $\nu = \sqrt{1-AK_2(K_2-1)/q}$.

In fact (see \(L action) if $\partial \Psi/\partial x^1 = 0$), the solutions in
\(bessel) are exact wavefunctions, satisfying $\L \Psi_0 = E\Psi_0$.

In the case where $\Psi(\rho,\xi^1)$ is not independent of $\xi^1$,
$\Psi(\rho,\xi^1)\not= \Psi_0(\rho,\xi^1)$. We can, in this case, solve \(psi
zero
2) asymptotically (i.e. for $\rho \mapsto 0$) to find the leading behavior of
wavefunctions near the $\rho =0$ singularity. \(psi zero 2) gives:
\beq
\Psi_0''+{3\over \rho}\Psi_0'
\sim -{CK_1^2\over q\rho^4}\Psi_0,\qquad
\rho \mapsto 0
\eeq\label{psi zero 3}

Solving \(psi zero 3) asymptotically we find:
\beq
\Psi_0(\rho) \sim e^{{+\atop -}\sqrt{{C\over -q}}{K_1\over \rho}},\qquad
\rho \mapsto 0
.\eeq

So that, finally:
\beq
\Psi_0(\rho,\xi^1) \sim e^{K_1\xi^1}
e^{{+\atop -}\sqrt{{C\over -q}}{K_1\over \rho}},\qquad
\rho \xi^1 \mapsto 0
\eeq
which shows that wavefunctions having an $\xi^1$ dependence ($K_1\not= 0$)
generically have an essential singularity at $\rho = 0$.

\sect{\bf 7. Some Further Directions}

We record here some ideas that should be interesting to pursue further.

Although Hill's equation
\beq
	\psi''+\hat q\psi=0
\eeq
is invariant under $\diff$, the eigenvalue problem
\beq
	\psi''+\hat q\psi=\lambda \psi
\eeq
is not invariant. This is because the LHS transforms as a density of weight
${3\over 2}$ while the RHS is of weight $-\half$. The determinant of Hill's
operator then, need not be diffeomorphism invariant: it could have an anomaly.
This anomaly can be calculated by the zeta--function method \[parker] and  we
 find  it to be zero. Thus the Hill determinant is a
diffeomorphism  invariant function of $q$ and can be written in terms of $\tr
M_q$. Assuming diffeomorphism invariance, we can show that,
\beq
	\det\;[-D_q]=\tr M_q-2.
\eeq
An argument for this would be note that if $D_q$ has no kernel, $q$ can be
transformed to a constant ${{\hat q}_0}$ by a diffeomoprhism (see below,
\[segal], and \[lazutkin]). The LHS can be evaluated to be
\beq
	\prod_{n\in Z}[n^2-{{\hat q}_0}]=(-{{\hat q}_0})
\prod_{n=1}^{\infty}n^4\prod_{n=1}^{\infty}[1-{{{\hat q}_0}\over n^2}]^2.
\eeq
The first infinite product is divergent but can be gievn a meaning by zeta
function regularization,
\beq
	\prod_{n=1}^{\infty}n^4=e^{-4\zeta'(0)}=(2\pi)^2,
\eeq where
\beq
	\zeta(s)=\sum_{n=1}^{\infty}{1\over n^s}
\eeq is the Riemann zeta function.

Now recall the standard formula
\beq
	{\sin z\over z}=\prod_{n=1}^{\infty}[1-{z^2\over n^2\pi^2}].
\eeq
With $z=\pi\surd {{\hat q}_0}$ this  can be used to evaluate the last factor,
so that
\beq
	\det\;[-D_{{{\hat q}_0}}]=-4 \sin^2[\pi\surd {{\hat q}_0}]=\cos 2\pi\surd
 {{\hat
q}_0}-1.
\eeq
We can solve the differential equation easily to determine the monodromy
operator  $M_{{{\hat q}_0}}$ and get
\beq
	\tr M_{{{\hat q}_0}}=2\cos 2\pi\surd {{\hat q}_0}.
\eeq
Thus we get the above result for constant $q$.  Through diffeomorphism
invariance, the identity follows for any $q$ for which $D_q$ is invertible,
because such a $q$ can transformed to a constant. If
$D_q$ is not invertible, both sides vanish and the identity is still true.

Another way to construct a diffeomorphism invariant from $q$ is to consider the
nonlinear eigenvalue problem
\beq
	\psi''+\hat q\psi=-\mu \psi^{-3},\quad \int \psi^{-2} dx=1
\eeq\label{eigen}
This nonlinear equation is invariant under $\diff$ since $\psi^{-3}$ is a
density of weight ${3\over 2}$. The normalization condition on $\psi$ also is
invariant.  If such  a $\psi$ exists that is smooth, periodic and positive
everywhere,  there is a diffeomorphism $\phi_q\in \diff$ which will transform
$q$ to a constant ${\hat q}_0$. This can be verified by putting,
\beq
	\psi(x)=(2\pi)^{-\half}\phi_q'^{-\half}(x),\quad
\hat q(x)=\hat q_0\phi_q'^2(x)+ S_{\phi_q}(x), \quad \mu=-2\pi{\hat q}_0.
\eeq
Conversely, whenever such a $\phi_q$ exists, we can solve the above nonlinear
eigenvalue problem. In Ref.\[segal] a sufficient condition for the existence
of $\phi_q$ is given: $M_q$ should lie on a one--parameter subgroup. This
happens whenever $D_q$ is invertible, for then, $M_q$ is diagonalizable. Thus
in
these situations $\psi$ will exist. Moreover, if $q_0$ so determined is not
equal to ${m^2\over 4}$ for $m\in Z$, the solution $\psi$ is unique, and
therefore $\phi_q$ is also determined uniquely modulo an additive constant
corresponding to a rotation. If $q_0={m^2\over 4}$  there are an additional
periodic solutions,
which can be easily determined in the co--ordinate system where $q=q_0$:
\beq
	\psi^2(x)=C+\sqrt{C^2-4\pi^2} \cos m(x-x_0),\quad \mu=-\pi^2m^2.
\eeq
Here $C$ and $x_0$ are constants of integration.

It would be interesting to give an independent proof of the existence of the
solution to the above nonlinear equation. It can be viewed as the
Euler--Lagrange equation of the variational problem
\beq
	\mu=\inf_\psi\big\{ \int[\psi'^2(x)-\hat q(x)\psi^2(x)]dx;\;  \int
|\psi|^{-2}(x)dx=1\big\}.
\eeq

This  problem is many ways analogous to the Yamabe problem \[yamabe]
 of differential geometry.  Recall that if $(M,g)$ is a compact $m$-dimensional
Riemannian manifold, ( $m\geq 3$)  the Yamabe variational problem is
\beq
	\mu=\inf\big\{\int [|d\psi|^2 +aR\psi^2]dV_g;\;\int |\psi|^r dV_g=1\big\}.
\eeq
Here
\beq
	a={m-2\over 4(m-1)},\quad r={2m\over m-2}
\eeq
and $dV_g$ is the volume measure defined by $g$ and $R$  the scalar curvature.
The variational equation is,
\beq
	\Delta \phi +a R\phi= -\mu\phi^{r-1}.
\eeq
Whenever such a $\phi$ exists and is positive, there is a new metric $\tilde
g=\phi^{r-2} g$ with constant scalar curvature.

The operator $\Delta+aR$ is the conformal Laplacian which maps scalar densities
of weight ${m\over 2}-1$ to those of weight ${m\over 2} +1$. The homogenous
equation
\beq
	[\Delta +aR]\psi=0
\eeq
is conformally invariant while the corresponding linear eigenvalue problem is
not.

The Yamabe problem is of course trivial when $m=1$: all Riemann metrics  are
then flat. ( When $m=2$ the  analogous equation is the Liouville equation).
Yet, our nonlinear eigenvalue problem \(eigen) is in many ways  a
one--dimensional limiting case of the Yamabe problem. Hill's operator
${d^2\over dx^2}+\hat q$ maps densities of weight $-\half$ to those of weight
${3\over 2}$, which is thus analogous to the conformal Laplacian, if we put
$m=1$ in the above formulae. Also, the Yamabe eigenvalue problem  becomes our
problem \(eigen) if we replace the conformal Laplacian by  Hill's operator and
put $m=1$. In this point of view, $q$ is analogous to the Ricci scalar rather
than metric and $\psi$ determines the transformation that reduces it to a
constant.

We have argued that for generic $q$, there is a unique $\psi$
that solves the eigenvalue problem. This allows us to define a new metric on
the space $Q$:
\beq
	||\delta q||^2=(2\pi)^3\int \delta q^2(x)\psi^6(x)dx=\int \delta
q^2(x)\phi_q'^{-3}(x)dx.
\eeq
This new metric is invariant under the full group ${\rm Diff}(S^1)$. This
metric is also  worth a detailed study. In particular, we
believe it will be of interest in understanding the geometry of the space of
one--dimensional potentials in quantum mechanics as well as quantum gravity on
a circle.

Although we have mainly discussed the generic case, the case where the isotropy
group of $q$ is smaller than $S^1$ ( when it cannot be transformed  to a
constant)  or larger ( when ${\hat q}_0={m^2\over 4}$ and the isotropy group is
three dimensional, the $m$--fold cover  $SL_2^m(R)$ of $PSL_2(R)$) are also of
interest. The analysis of the latter case is quite straightforward: there is a
two parameter family of solutions to the eigenvalue problem \(eigen)
as described above.  If $C>2\pi$ this solution  is positive and will
define a diffeomorphism. There is a three parameter family of diffeomorphisms
that will reduce such a $q$ to a constant, $q_0={m^2\over 4}$. The
corresponding
co--adjoint orbit will be $\diff/SL_2^m(R)$.
The case where the isotropy group does not contain rotations, $S^1$, is more
complicated; their study will involve a detailed understanding of the Mathieu
equation.

The Hill's determinant can be thought of in terms of a Gaussian path integral:
\beq
 Z_0[q]=\int {\cal D}\psi e^{-S[\psi,q]}=\det\;^{-\half}[-D_q]
\eeq
where
\beq
	S_0[\psi,q]=\half\int_0^{2\pi}[\psi'^2-\hat q(x)\psi^2]dx.
\eeq
The path integral is over all configurations of period $2\pi$.
This is a free one--dimensional quantum  field theory. We can ask if there is
 way
a way to add interactions without destroying diffeomorphism invariance. The
answer is,
\beq
	Z_\mu[q]=\int {\cal D}\psi e^{-S_\mu[\psi,q]}
\eeq
where,
\beq
		S_\mu[\psi,q]=\half\int_0^{2\pi}[\psi'^2-\hat q(x)\psi^2-\mu\psi^{-2}]dx.
\eeq
  If $\mu=0$, this reduces to  the  expression of the Hill's determinant.
The classical equation of this field theory is the nonlinear eigenvalue problem
we discussed earlier.
It is more difficult to see if this generalization of the Hill's determinant is
diffeomorphism invariant directly, since there is no analogue for the zeta
function argument in the full interacting theory.  One approach would be a
perturbation theory in $\mu$. But, if we assume that $Z_\mu(q)$  is invariant,
 it is
possible to evaluate this path integral in the case of  generic $q$.

Then we can transform to co--ordinates where $\hat q={\hat q}_0$. Then this
becomes the action of a point particle in a harmonic oscillator plus inverse
square potential. This one dimensional quantum  field theory has been studied
by Parisi and Zirilli \[parisi]. (See also Gupta and Rajeev\[rengrp].)
 In fact, by viewing this in the canonical language, where $\psi$ would be the
position variable and $x$ would be time,
\beq
	Z_\mu[{\hat q}_0]=\tr e^{-2\pi H_\mu}
\eeq
where  the hamiltonian is
\beq
H_{\mu}=-\half[{d^2\over d\psi^2}+{\hat q}_0\psi^2  +\mu \psi^{-2}].
\eeq
The eigenvalues of this hamiltonian are ( after correcting an algebraic error
in
  Ref.\[parisi])
\beq
	E_n=\surd(-{\hat q}_0)[n+1-\half\surd(1-4\mu)]
\eeq
when $q_0<0,\mu<{1\over 4}$. In this range we can get the partition function,
\beq
Z_{\mu}(q)=Z_\mu(q_0)={1\over 2\sinh[\pi\surd(-q_0)]}
e^{\pi(\sqrt{1-4\mu}-1)\sqrt{-q_0}}.
\eeq
This agrees with $\det\;^{-\half}[-D_q]$ when $\mu=0$.
For  values of $q_0$ and $\mu$ at which the above argument does not apply, we
 can define $Z_\mu(q_0)$  by analytic continuation. Note that
there is a branch point at $\mu={1\over 4}$, which is the critical value
observed in  Ref. \[rengrp]. When $\mu>{1\over 4}$, we must perform a
renormalization in order to get a sensible ground state, as described in
 Ref.\[rengrp]. This would be
necessary to get a meaningful value of $Z_\mu$ as well.

{\bf Acknowledgements}

We thank K. Gupta, S. Guruswamy, T. Turgut, and P. Vitale for discussions on
this topic. This work has been supported in part by the US Department of
Energy, Grant No. DE-FG02-91ER40865.

\vfill\eject

\noindent{\bf References}\hfill\break

\noindent\wittena. Witten E 1990 {\it Cambridge 1990, Proceedings, Surveys in
Differential Geometry} 243

\noindent\hajicek. Hajicek P 1992 {\it Class. Quantum Grav.} {\bf 9} 2249

\noindent\rajeeva. Rajeev S G 1982 {\it Phys. Lett.} {\bf 113B} 146

\noindent\rajeev. Rajeev S G 1988 {\it Phys. Lett.} {\bf B212} 203

\noindent\dewitt. Dewitt B S 1967 {\it Phys. Rev.} {\bf 162} 1195

\noindent\einstein. Einstein A 1956 {\it The Meaning of Relativity} (Princeton:
Princeton University Press)

\noindent\lanczos. Lanczos C 1970 {\it The Variational Principles of Mechanics}
(New York: Dover)

\noindent\whittaker. Whittaker E T and Watson G N 1927 {\it A Course of Modern
Analysis} (Cambridge: Cambridge University Press)

\noindent\segal. Segal G 1981 {\it Comm. Math. Phys.} {\bf 80} 301

\noindent\eisenhart. Eisenhart L 1949 {\it Riemannian Geometry} (Princeton:
Princeton University Press)

\noindent\god. Goddard P and Olive D 1986 {\it Int. J. Mod. Phys.} {\bf A1} 303

\noindent\bowicka. Bowick M J and Rajeev S G 1987 {\it Nucl. Phys.} {\bf B293}
348

\noindent\bowickb. Bowick M J and Rajeev S G 1987 {\it Phys. Rev. Lett.} {\bf
58} 535

\noindent\witten. Witten E 1988 {\it Comm. Math. Phys.} {\bf 114} 1

\noindent\bieber. Thurston W P 1986 in {\it The Bieberbach Conjecture:
Proceedings of The Symposium on the Occasion of the Proof} ed A Baernstein II,
D Drasin, P Duren and A Marden (Providence:
American Math. Society)

\noindent\scherer. Scherer W 1988 {\it Thesis} University of Rochester

\noindent\polyakov. Polyakov A M 1987 {\it Gauge Fields and Strings} (Chur:
Harwood Academic)

\noindent\rengrp. Gupta K S and Rajeev S G 1993 {\it Phys. Rev.} {\bf D48} 5940

\noindent\parker. Parker T and Rosenberg S 1987 {\it J. Diff. Geo.} {\bf 25}
199

\noindent\lazutkin. Lazutkin V F and Pankratova T F 1975 {\it Funct. Anal.
Applic.} {\bf 9} 306

\noindent\yamabe. Aubin T 1982 {\it Nonlinear Analysis on Manifolds,
Monge-Amp\`ere Equations} (New York: Springer-Verlag)

\noindent\parisi. Parisi G and Zirilli F 1973 {\it J. Math. Phys} {\bf 14} 243

\bye